\documentclass[AMA,STIX1COL]{template}
\usepackage{moreverb}
\usepackage{adjustbox}
\usepackage{bm}

\usepackage[usestackEOL]{stackengine}
\usepackage{float}
\usepackage{authblk}        % for affiliations
\newcommand\BibTeX{{\rmfamily B\kern-.05em \textsc{i\kern-.025em b}\kern-.08em
T\kern-.1667em\lower.7ex\hbox{E}\kern-.125emX}}

\makeatletter
\renewcommand\@maketitle{
   \vspace*{-2.5em} % Pull content closer to the top of the page
  \noindent\rule{\textwidth}{0.4pt}  % Top horizontal line
  \vspace{1em}   % Space after the line (adjust as needed)
  \begin{center}
    {\LARGE\bfseries \@title \par}
    \vskip 1.5em
    {\@author}
    \vskip 1.5em
    \textbf{ *Corresponding Author:} Guangyu Tong, PhD \\
    \textbf{Email:} \texttt{guangyu.tong@yale.edu}
    \vskip 1em
    {\large \@date}
  \end{center}
}
\makeatother

\title{Bayesian inference for cluster-randomized trials with multivariate
outcomes subject to both truncation by death and missingness}

\author[1,2,3,4]{Guangyu Tong*}

\author[5]{Chenxi Li}

\author[1]{Eric Velazquez}

\author[6,7]{Michael O. Harhay}

\author[1,2,4]{Fan Li}

\affil[1]{Department of Internal Medicine, Section of Cardiovascular Medicine, Yale School of Medicine, New Haven, CT, USA}

\affil[2]{Department of Biostatistics, Yale School of Public Health, New Haven, Connecticut, USA}

\affil[3]{Cardiovascular Medicine Analytics Center, Yale School of Medicine, New Haven, Connecticut, USA}

\affil[4]{Center for Methods in Implementation and Prevention Science, Yale University, New Haven, Connecticut, USA}

\affil[5]{Department of Biostatistics \& Bioinformatics, Duke School of Medicine, Durham, North Carolina, USA}

\affil[6]{Clinical Trials Methods and Outcomes Lab, Palliative and Advanced Illness Research (PAIR) Center, Perelman School of Medicine, University of Pennsylvania, Philadelphia, Pennsylvania, USA}

\affil[7]{Department of Biostatistics, Epidemiology and Informatics, Perelman School of Medicine, University of Pennsylvania, Philadelphia, Pennsylvania, USA}

\begin{document}

\maketitle
\thispagestyle{empty} 
\vspace{0.5em}
\begin{center}
\textbf{Abstract}
\end{center}

\begin{center}
\begin{minipage}{0.85\textwidth} 
Cluster-randomized trials (CRTs) on fragile populations frequently encounter complex attrition problems where the reasons for missing outcomes can be heterogeneous, with participants who are known alive, known to have died, or with unknown survival status, and with complex and distinct missing data mechanisms for each group. Although existing methods have been developed to address death truncation in CRTs, no existing methods can jointly accommodate participants who drop out for reasons unrelated to mortality or serious illnesses, or those with an unknown survival status. This paper proposes a Bayesian framework for estimating survivor average causal effects in CRTs while accounting for different types of missingness. Our approach uses a multivariate outcome that jointly estimates the causal effects, and in the posterior estimates, we distinguish the individual-level and the cluster-level survivor average causal effect. We perform simulation studies to evaluate the performance of our model and found low bias and high coverage on key parameters across several different scenarios. We use data from a geriatric CRT to illustrate the use of our model. Although our illustration focuses on the case of a bivariate continuous outcome, our model is straightforwardly extended to accommodate more than two endpoints as well as other types of endpoints (e.g., binary). Thus, this work provides a general modeling framework for handling complex missingness in CRTs and can be applied to a wide range of settings with aging and palliative care populations.
\end{minipage}
\end{center}

\vspace{1em}


\section{Introduction}\label{sec:intro}
Cluster-randomized trials (CRTs) are a popular class of trial designs where the intervention under test is allocated to entire groups or clusters rather than individuals themselves.\citep{Murray1998,Turner2017a,Turner2017b} Many CRTs that evaluate non-mortality outcomes such as quality of life (QoL) can encounter significant attrition before the outcomes are fully collected or measured. \citep{colantuoni2018statistical} In real-world settings, the reasons for missing outcomes can be diverse. \citep{fiero2016statistical} In vulnerable study populations like the elderly, or patients in the critical care settings, many participants can have outcomes missing due to death or deteriorated health. \citep{rubin2006causal} As a result, by the time of the outcome assessment, the non-mortality outcomes will be left missing and at most ambiguously defined. Participants can also have missing outcomes for other reasons, such as temporarily relocating or transferring to other healthcare facilities. Participants in this situation are likely to be alive but just miss the final visit for outcome assessment. In addition, participants can also be lost to follow up, migrate into other geographical areas, or other undocumented reasons that are distinctly different than death. If these reasons are not related to health deterioration or death, both mortality status and non-mortality outcomes may be missing (but for different reasons) at the time of assessment. 

Under the circumstance of outcomes truncation by death, complete case analysis can suffer from the survivor bias (i.e., observed and unobserved survivor characteristics being systematically different between the treatment and control groups) when estimating the intervention effect on the non-mortality outcomes. \citep{colantuoni2018statistical} Imputation-based approaches or inverse probability weighting are popular solutions for missing at random (MAR) scenarios, \citep{tong2022missing,turner2020properties} but they may not be appropriate to address the missingness caused by death truncation because non-mortality outcomes are conceptually defined with ambiguity among those who die. \citep{fielding2008simple} That is, these approaches provide the 'hypothetical' effect estimates in a setting where mortality didn't occur. On the other hand, composite outcomes can sometimes be employed to provide subject valuations for the non-mortality outcomes of dead individuals, but this approach also raises conceptual and interpretive issues as the survivor and decedent population is combined with a subjective rule. \citep{lin2017placement,Harhay2019} For example, components might have different levels of severity or clinical importance, and thus, the precise causal interpretation of results can also be challenging. \citep{Freemantle2003Composite} 
%More importantly, all solutions using complete-case analysis, imputation, and composite outcomes cannot produce estimates with a clear causal interpretation. 
Finally, as increasingly emphasized in the randomized clinical trials literature (e.g., Wang et al.\citep{wang2024model} and Kahan et al. \citep{kahan2022estimands}), a clearly defined causal estimand is especially important for CRTs. In particular, the individual-average causal effect and the cluster-average causal effect can capture fundamentally different aspects of an intervention and address different levels of hypotheses, but variations of such estimands in the presence of outcomes truncated by death have not been previously proposed. 

To clearly define the causal estimand, previous works have addressed the outcome truncation due to an intermediate outcome, such as death, under the principal stratification framework. \citep{Frangakis2002, Zhang2009,Ding2011Identification,Zhang2003estimation,YangSmall2016,Tchetgen2014} Based on the pre-treatment characteristics, an important subpopulation of participants (principal stratum) can be defined as the \emph{always-survivors}--the participants who will survive regardless of treatment assignment. Under the potential outcomes framework \citep{rubin1974estimating, neyman1934two}, only participants in this stratum have both potential outcomes well-defined. Therefore, the estimand of interest can be defined as the average contrast between potential outcomes among the always-survivors stratum, namely, the survivor average causal effect (SACE).\citep{Zhang2003estimation,rubin2006causal} %The SACE avoids survivor bias and estimates the intervention effect excluding those whose has one as two counterfactual survivor statuses as non-survivors. 
Because the always-survivors stratum membership is not fully observed, Bayesian inference provides an effective tool to jointly identify the stratum membership in a principle strata model ($S$ model) and estimate the SACE in an outcome regression model ($Y$ model) conditional on the strata membership. In the context of CRTs, individuals within the same cluster tend to have similar outcomes, due to a positive intracluster correlation coefficient (ICC).\citep{Turner2017a} Failing to accommodate this correlation in the analysis can lead to inflated type I error rates. With a single non-mortality outcome, Tong et al. \cite{tong2023bayesian} studied a Bayesian model for estimating the SACE while accounting for the cluster-induced correlation in both the $S$ model and the $Y$ model. They found that the bias and coverage of the SACE are not as sensitive to the correlation specification in the $S$ model. In the same context, Wang et al.\cite{Wang2024EM} proposed a maximum likelihood approach for the estimation of the SACE using the Expectation-Maximization algorithm. They found that a weakly predictive $S$ model can lead to a larger variance of the SACE estimator. 

Despite these previous works, no methods have been developed in CRTs to simultaneously address truncation by death and dropout due to reasons other than mortality or serious illnesses, which are common in real-world settings. Excluding these dropout participants--who have both unknown survival status and missing outcomes--can lead to potentially biased results as well as power loss. The multilevel multiple imputation approach\cite{turner2020properties} holds promise to address the participant attrition for those who are likely alive, but has not been integrated with methods that also address outcome truncation by death. In addition, little attention has been paid to the joint estimation of multiple outcomes for SACE in CRTs, while multiple outcomes are increasingly common as explained in recent systematic reviews.\cite{taljaard2022methodological,nevins2022review} For instance, QoL assessment can include separate physical and mental health measures, and a multivariate outcome model for both outcomes can lead to efficiency gain in CRTs when the cluster size is variable,\cite{yang2023power} even without missing or truncated outcomes. In the presence of truncation by death, as shown by Mercatnti et al. \cite{mercatanti2015improving} and Mattei et al. \cite{mattei2013exploiting} for independent data, including a second outcome can effectively sharpen the inference for the parameters in the $S$ model, and the improvement is more apparent when the distance between principle strata components increases.

In this article, we simultaneously address missing and truncated bivariate outcomes in CRTs by proposing a Bayesian model for estimating SACE in CRTs. Under the potential outcomes framework, we extend the recent estimands discussion in Kahan et al.\cite{kahan2022estimands} and formally define the SACE estimand both as an individual average as well as a cluster average. We consider an empirical identification strategy for the strata of always survivors by imposing a mixture of bivariate linear mixed models given the counterfactual survivor status. \citep{tong2023bayesian, Wang2024EM} For participants with both missing survival status and outcomes, we consider a nested missing at random (nested MAR) assumption, that is, (i) given baseline covariates, the missingness of the mortality status is independent of the potential outcomes and potential survivor status, and (ii) given the baseline covariates and that the patient has survived, the missingness of the outcome is independent of the potential outcomes. Our model can be easily extended to accommodate other types of endpoints or a multivariate outcome with three or more dimensions.

The rest of our paper is organized as follows. Section 2 outlines the principal stratification framework and introduces the structural assumptions for causal inference in CRTs. Section 3 specifies the $S$ model, bivariate linear mixed $Y$ model and the likelihood in the context of CRTs. We also define the ICC parameters that are meaningful for a bivariate continuous outcome and explain the model specification and strategies for non-continuous outcomes. Section 4 describes the Bayesian inference with the prior specification and the pseudo algorithm for posterior inference. Section 5 presents the design and results of simulation studies that are used to investigate the performance of our proposed method. Section 6 applies our method to data from the Whole Systems Demonstrator (WSD) Telecare Questionnaire Study, a pragmatic CRT comparing participant-reported QoL outcomes between a telecare intervention and usual care. \citep{Hirani2013ageing}  Section 7 concludes with a discussion of the use of our methods in practice.

\section{Methods}\label{sec:methods}
\subsection{Notation and setup}\label{sec:notations}
We consider a two-arm CRT with $n$ clusters and each with $N_i$ individuals, where $i = 1\dots n$. Let $Z_{i}$ represent the binary cluster-level treatment assignment, where $Z_{i}=1$ indicates treatment and $Z_{i}=0$ indicates control. We denote $\boldsymbol{X}_{ij}$ as the $p$-vector of baseline covariates for $j$th individual in $i$th cluster (this vector can include an intercept, individual-level as well as cluster-level covariates and cluster size $N_i$ for notation brevity). We are interested in jointly estimating the causal effect of $Z_i$ on a multivariate outcome with $K$ dimensions, $\boldsymbol{Y}_{ij}=\left(Y^1_{ij},\ldots, Y^K_{ij}\right)^T$.
We subscribe to the potential outcomes framework and define the causal effect as the expected difference between the two potential outcomes of a common population under treatment and control. \citep{rubin1974estimating}  We define $\{\boldsymbol{Y}_{ij}(1), \boldsymbol{Y}_{ij}(0)\}$ as the potential outcomes for each individual under treatment and control measured at the end of the study; $\{S_{ij}(1)$, $S_{ij}(0)\}$ as the potential survival status of an individual under treatment and control at the time of final outcome assessment, with $S_{ij}=1$ indicating survival and $S_{ij}=0$ indicating death. Only one of the potential outcomes and one of the potential survivor status under the treatment assignment $Z_i=z$ can be observed. For those with $S_{ij}(z)=0$ (death), the final outcome $\boldsymbol{Y}_{ij}$ is not well-defined and for them we set $\boldsymbol{Y}_{ij}(z)=\star^{K\times 1}$ extending the notation in Zhang et al.\cite{Zhang2003estimation} 

We pursue the (i) Stable Unit Treatment Value Assumption (SUTVA) so that $\boldsymbol{Y}_{ij}=Z\boldsymbol{Y}_{ij}(1)+(1-Z)\boldsymbol{Y}_{ij}(0)$, which requires no difference in the versions of treatment the trial participants receive and no between-participant interference. We further assume that (ii) the treatment is randomized at the cluster level and independent of both the potential survivor status and counterfactual outcomes of all individuals in each cluster; this can be expressed as $Z_i\,\bot\, \{S_{ij}(1),S_{ij}(0),\boldsymbol{Y}_{ij}(1),\boldsymbol{Y}_{ij}(0), j=1,\ldots,N_i\}$.

% Following Wang et al.,\cite{wang2024model} we can define two different estimands for the treatment effect in CRTs: (i) the individual average causal effect as $\boldsymbol{\delta}_{I}=E[\mu_I(\boldsymbol{Y}_{ij}(1))-\mu_I(\boldsymbol{Y}_{ij}(0))]$, and (ii) the cluster average causal effect as $\boldsymbol{\delta}_{C}=E[\mu_C(\boldsymbol{Y}_{ij}(1))-\mu_C(\boldsymbol{Y}_{ij}(0))]$. Here, $\mu_I$ and $\mu_C$ are mean functions for the individual-level and cluster-level averages. Under death truncation, these quantities can only be clearly defined among individuals with both potential outcomes defined and measured ($\{S_{ij}(1)$, $S_{ij}(0)\} = \{1,1\}$). This subset of individuals will be the group of interest with $\boldsymbol{\delta}_{I}$ and $\boldsymbol{\delta}_{C}$ meaningfully defined. For those with either or both potential outcomes as $\star^{K\times 1}$, $\boldsymbol{\delta}_{I}$ and $\boldsymbol{\delta}_{C}$ cannot be expressed meaningfully. 

\subsection{Principal stratification and estimands}\label{sec:ps}
To identify the subset of individuals who can always survive under treatment and control, we subscribe to the framework of principal stratification that uses the potential values of a post-treatment intermediate variable to define a cross-classification of participants into different subpopulations. We can then compare the potential outcomes to define meaningful causal estimands in each subpopulation, or principal stratum. Given a binary treatment assignment $Z_i$ and a binary survival status $S_{ij}$, individuals can be classified into four principal strata based on the joint values of $\{S_{ij}(1), S_{ij}(0)\}$, which include:
\begin{itemize}
    \item \emph{always-survivors} ($S_{ij}(1)=1, S_{ij}(0)=1$): individuals who survive regardless of the treatment assignment.
    \item \emph{protected} ($S_{ij}(1)=1, S_{ij}(0)=0$): individuals who survive only under treatment.
    \item \emph{harmed} ($S_{ij}(1)=0, S_{ij}(0)=1$): individuals who survive only under control.
    \item \emph{never-survivors} ($S_{ij}(1)=0, S_{ij}(0)=0$): individuals who will not survive regardless of the treatment assignment.
\end{itemize}

Here, we make an additional assumption of monotonicity such that $S_{ij}(1)\geq S_{ij}(0)$, \citep{angrist1996identification, Zhang2003estimation, tong2023bayesian} so that the harmed strata is ruled out. This assumption rules out participants who would only survival under the control condition but would die under the test intervention. It is usually plausible in trials where interventions are carefully evaluated in pilot studies and perceived to be beneficial for participant survival. However, monotonicity may be violated when, for example, in a trial comparing two active treatments with unknown relative benefits. Sensitivity analysis methods have also been proposed to evaluate the impact when monotonicity is violated, and are a subject for future research in CRTs.\citep{DingLu2017,Chiba2011aje,shepherd2008does} Note that in scenarios where the proportion of the harmed strata is close to zero, incorporating the harmed strata into the modeling process may lead to greater computational challenges and numerical instability, in which case enforcing monotonicity may also be a practical consideration for point estimation. 

Under monotonicity, some participants will have known principal strata membership based on their observed survivor status. Participants who are observed as survivors under the control condition will always survive if they were assigned to treatment, and will be identified as always-survivors. Participants who are observed as non-survivors under the treatment condition will not survive if they were assigned to control, and will be identified as never-survivors. However, participants who are observed as non-survivors under the control condition can be from either the protected or never-survivor stratum, whereas participants who are observed as survivors under the treatment condition can be from either the always-survivor or protected stratum. 

To simplify the notations for the principal strata membership, let $G_{ij}$ indicate principal strata membership and $\boldsymbol{\pi}_{ij}$ be the vector of probabilities of strata membership, where $G_{ij}=00$ indicates the never-survivors stratum; $G_{ij}=10$ indicates the protected stratum; $G_{ij}=01$ indicates the harmed stratum; $G_{ij}=11$ indicates the always-survivors stratum. Under monotonicity, we assumed away the harmed strata, and $\boldsymbol{\pi}_{ij}$ can, therefore, be written as a vector with three elements summing to one. Since the pair of potential outcomes is only well-defined among the always-survivors stratum, we can define the following two causal estimands in CRTs for studying the treatment effect on the non-mortality outcomes. Extending the estimands discussion in Kahan et al.\cite{kahan2023estimands,kahan2024demystifying} and Wang et al.,\cite{wang2024model}, the survivor individual-average causal effect (SIACE) can be defined as 
$$\boldsymbol{\delta}_{I}=\bm{\mu}_I(1)-\bm{\mu}_I(0)=\left(\begin{matrix}
\mu_I^1(1)\\
\vdots\\
\mu_I^K(1)
\end{matrix}\right)-\left(\begin{matrix}
\mu_I^1(0)\\
\vdots\\
\mu_I^K(0)
\end{matrix}\right)$$
where $$\mu_I^k(z)=\frac{E\left[\sum_{j=1}^{n_i}I(G_{ij}=11) {Y}_{ij} (z)\right]}{E\left[\sum_{j=1}^{n_i}I(G_{ij}=11)\right]},~~~k=1,\ldots,K.$$
On the other hand, the survivor cluster-average causal effect (SCACE) can be defined as 
$$\boldsymbol{\delta}_{C}=\bm{\mu}_C(1)-\bm{\mu}_C(0)=\left(\begin{matrix}
\mu_C^1(1)\\
\vdots\\
\mu_C^K(1)
\end{matrix}\right)-\left(\begin{matrix}
\mu_C^1(0)\\
\vdots\\
\mu_C^K(0)
\end{matrix}\right),$$
where $$\mu_C^k(z)=E\left[\frac{{\sum_{j=1}^{n_i} I(G_{ij}=11) {Y}_{ij} (z)}}{\sum_{j=1}^{n_i}I(G_{ij}=11)}\right],~~~k=1,\ldots,K.$$

It is worth noting that these two estimands are conceptually different, because SIACE addresses an individual-level hypothesis among the population of all participants, whereas SCACE addresses a cluster-level hypothesis among the population of clusters.\cite{kahan2022estimands} Apart from this conceptual difference, the key factor that drives the numerical difference between the two estimands is whether the number of always-survivors in each cluster is informative of the intervention effect. In other words, the difference between SCACE and SIACE hinges on the presence of \emph{informative cluster size}, which arises when there is a non-zero marginal correlation between the number of always-survivors within each cluster and the non-mortality outcomes. Finally, because the observed survivors are a mixture of always-survivors and the protected, identification of the mixture membership will be a crucial step to generate estimates for both the SIACE and SCACE estimands. 

\subsection{Missing outcomes}\label{sec:missingnotduetodeath}

Let $R_{ij}^S$ be the missingness indicator for the survival status and $R_{ij}^Y$ be the missingness indicator for the final non-mortality outcome, with 0 indicating missingness and 1 indicating the information is fully observed. We assume that all components of a multivariate outcome are either all missing or observed such that a scalar $R_{ij}^Y$ can represent missingness without ambiguity. Participants who drop out due to non-health-related reasons, such as temporarily moving out of a trial site, have a high likelihood of being alive with $S_{ij}=1$. They just have missing non-mortality outcomes (e.g., QoL outcomes), which can be handled with an outcome imputation model leveraging the covariate information under the missing at random assumption.\citep{tong2022missing}. However, for participants who drop out for unclear reasons, both their survival status and final outcomes are not observed. Here, we write the complete data with all variables for each individual as $\boldsymbol{W}_{ij}=\{\boldsymbol{Y}_{ij}(1), \boldsymbol{Y}_{ij}(0), S_{ij}(1), S_{ij}(0), \boldsymbol{X}_i\}$, and different fractions of the complete data $\boldsymbol{O}_{ij}$ may be observed in each individual. We summarize the observed data vectors under several different scenarios in Table \ref{obs}. In short, we classify the observed data into four different patterns: (i) the individual has survived until the time of final outcome assessment and the final non-mortality outcome is fully observed; (ii) the individual final outcome is truncated by death; (iii) the individual has survived until the time of final outcome assessment but the outcome is not available due to loss to follow-up with reasons other than death; (iv) the individual has lost to follow-up before the time of outcome assessment such that the vital status is not even available. 

\begin{table}[]
\caption {Missingness patterns and observed survival status and non-mortality outcomes}
\label{obs} 
\centering
\begin{tabular}{@{}ll@{}}
\toprule
Missingness patterns           & Observed data                                         \\ \midrule
No truncation by death and no missing outcomes & $\boldsymbol{O}_{ij}=\{\boldsymbol{Y}_{ij}, Z_i, \boldsymbol{X}_{ij}, S_{ij}=1, R_{ij}^S=1, R_{ij}^Y=1\}$                            \\
Truncation by death only                           & $\boldsymbol{O}_{ij}=\{\star^{K\times 1}, Z_i,  \boldsymbol{X}_{ij}, S_{ij}=0, R_{ij}^S=1, R_{ij}^Y=1\}$\\
Survived but with missing non-mortality outcomes      & $\boldsymbol{O}_{ij}=\{Z_i, \boldsymbol{X}_{ij}, S_{ij}=1, R_{ij}^S=1, R_{ij}^Y=0\}$                                \\
Missing both survival status and non-mortality outcomes    & $\boldsymbol{O}_{ij}=\{Z_i, \boldsymbol{X}_{ij}, R_{ij}^S=0\}$                           \\ \bottomrule
\end{tabular}
\end{table}

To address missingness related to incomplete measurements of outcomes and survival status, we make the nested MAR assumption, stated as
    $$\text{(A1)}~~R_{ij}^S \perp \left\{S_{ij}(1), S_{ij}(0), \boldsymbol{Y}_{ij}(1),\boldsymbol{Y}_{ij}(0); j=1,\ldots,N_i\right\}| \{Z_i=z,\boldsymbol{X}_{ij}; j=1,\ldots,N_i\};$$
    $$\text{(A2)}~~R_{ij}^Y \perp \{\boldsymbol{Y}_{ij}(z);j=1,\ldots,N_i\}|\{Z_i=z, \boldsymbol{X}_{ij}, S_{ij}(z)=1, R_{ij}^S =1,j=1,\ldots,N_i\}$$
In the first layer of the assumption, the missingness of the survivor status is independent of the counterfactual survivor status, the counterfactual outcomes, and the treatment assignment, conditional on the observed covariates within each cluster. In the second layer, among observed survivors, the missingness of the outcome is independent of the potential outcomes conditional on the covariates, treatment assignment, and the principal strata membership (i.e., either always survivors or protected) within each cluster. Under this assumption, the missing data mechanism is ignorable (hence not dependent on the unobserved information), so that the unbiased estimation of SIACE and SCACE is possible based on the observed data. Note that, the conditioned covariates in the two layers of the NMAR assumption do not necessarily need to be the same. For the simplicity of notations, we use the same covariate vectors to represent the baseline covariate information that can make the missingness ignorable. The principal strata membership model and stratum-specific outcome models will then serve as imputation models for participants with missing survival status and those with missing outcomes not due to death truncation. This can be viewed as an extension of the multilevel imputation model of CRTs in Turner et al. \cite{turner2020properties} Our approach also extends Bia et al., \cite{Bia2022} which assumes that an intermediate variable and an outcome are either both missing or both observed in an non-clustered observational study. With the NMAR assumption, we can additionally accommodate the scenarios where survival status is observed but study outcomes are missing.

\section{Principal stratification and model specifications}\label{sec:modeling}
\subsection{Likelihood}\label{sec:likelihood}

We consider a Bayesian joint modeling approach for the identification of the principal strata membership and the estimation of SIACE and SCACE. We can classify participants into four categories based on the observed treatment assignments and survival status at the time of outcome assessment:

\begin{itemize}
  \item $O(1,1)=\{(i,j)|Z_i=1,S_{ij}=1\}$, participants from the treated clusters and survived;
    \item $O(1,0)=\{(i,j)|Z_i=1,S_{ij}=0\}$, participants from the treated clusters and did not survive;
    \item $O(0,1)=\{(i,j)|Z_i=0,S_{ij}=1\}$, participants from the control clusters and survived;
    \item $O(0,0)=\{(i,j)|Z_i=0,S_{ij}=0\}$, participants from the control clusters and did not survive.
\end{itemize}

Under monotonicity, the stratum memberships for participants in $O(1,0)$ and $O(0,1)$ can be fully determined, whereas the stratum memberships for participants in $O(1,1)$ and $O(0,1)$ involve a mixture and cannot be directly inferred without additional modeling assumptions. We use $\boldsymbol{S}^{obs}$ to denote the collection of observed survival status for all individuals and $\boldsymbol{Y}^{obs}$ to denote the collection of observed non-mortality outcomes for all individuals. Let $\boldsymbol{\theta}$ denote the global parameters, and the principal strata membership model can be expressed as $p_{ij,g}=P(G_{ij}=g|\boldsymbol{X}_{ij},\boldsymbol{\theta})$. Let $f_{g,z}$ be a generic density function for the outcome, where $f_{ij,g,z}=P(\boldsymbol{Y}_{ij}(z)|G_{ij}=g,\boldsymbol{X}_{ij}\boldsymbol{\theta})$, for $g=00,10,11$ and $z=0, 1$. The likelihood function is then given by
\begin{align*}
L(\boldsymbol{Y}^{obs},\boldsymbol{S}^{obs},\boldsymbol{X},Z|\boldsymbol{\theta})=& \prod_{(i,j)\in \{Z_i=1,S_{ij}=1\}}\left\{p_{ij,11}f_{ij,11,1}+p_{ij,10}f_{ij,10,1}\right\}
 \times \prod_{(i,j)\in \{Z_i=1,S_{ij}=0\}}p_{ij,00} 
\times \prod_{(i,j)\in \{Z_i=0,S_{ij}=1\}}p_{ij,11}f_{ij,11,0}\\
& \times \prod_{(i,j)\in \{Z_i=0,S_{ij}=0\}}\left\{p_{ij,10}+ p_{ij,00}\right\}.
\end{align*}
In what follows, we assume the global parameter $\boldsymbol{\theta}$ can be split into two distinct sets of parameters---one set for parameterizing the principal strata membership model, and the other set for the outcome model. We will assume these two sets of parameters are \emph{a priori} independent to facilitate Bayesian inference.

\subsection{Model for principal strata membership}\label{sec:psmodel}
To account for the multilevel data structure in CRTs, we consider a nested Probit random-effects model for the principal strata membership.\citep{Frangakis2002NonCompliance} Under monotonicity, there are three strata---the always-survivors $G_{ij}=11$, the protected $G_{ij}=10$, and the never-survivors $G_{ij}=00$. The principal strata memberships are known for the deceased in the treatment clusters and for the survivors in the control clusters, but they remain unknown for the remaining participants. We thus leverage baseline covariates to identify the principal strata for those participants. 

The first layer of the nested Probit random-effects model determines whether an individual belongs to the never-survivor stratum ($G_{ij}=00$), while the second layer of the model determines whether an individual belongs to the protected stratum ($G_{ij}=10$). We define $\boldsymbol{\beta}$ and $\boldsymbol{\gamma}$ as the regression coefficients for the first and second layers and define $\chi_i$ as the cluster-level random intercept that follow a normal distribution with mean $0$ and variance $\phi^2$. The nested Probit random-effects model can be expressed as,
$$p_{ij,00}=1-\Psi\left(\boldsymbol{X}^T_{ij}\boldsymbol{\beta}+\chi_i\right)$$
$$p_{ij,10}=(1-p_{ij,00})\left(1-\Psi(\boldsymbol{X}^T_{ij}\boldsymbol{\gamma}+\chi_i)\right)$$
$$p_{ij,11}=1-p_{ij,00}-p_{ij,10}$$
Here, $\Psi$ is a cumulative distribution function of a standard normal distribution. Given that the always-survivors are usually the largest stratum, the above model specification helps stabilize the estimation. To assist Bayesian inference, we further pursue an equivalentlatent variable specification, through which the model can be further described as,
$$Q_{ij}|\boldsymbol{X}_{ij},\boldsymbol{\beta},\chi_i\sim N\left(\boldsymbol{X}^T_{ij}\boldsymbol{\beta}+\chi_i,1\right)$$
$$W_{ij}|G_{ij}\neq 00,\boldsymbol{X}_{ij},\boldsymbol{\gamma},\chi_i\sim N\left(\boldsymbol{X}^T_{ij}\boldsymbol{\gamma}+\chi_i,1\right),$$
and the principal strata membership will be determined by the model via
$$G_{ij}
    =\begin{cases}
      00 & \text{if $Q_{ij}> 0$, }\\
      10\ \text{or} \ 11 & \text{otherwise}
    \end{cases} \text{and} \ 
    G_{ij}
    =\begin{cases}
      10 & \text{if $W_{ij}> 0$, }\\
      11 & \text{otherwise}
    \end{cases}
    $$
%A multinomial logistic regression model can serve as an alternative model, and examples can be found in Tong et al.\cite{tong2023bayesian} and Wang et al. \cite{Wang2024EM} One advantage of using a nested Probit model is the latent variable specification that can help facilitate the model estimation.

\subsection{Model for bivariate continuous outcomes}\label{sec:continuous}
Without loss of generality but matching our application study in Section \ref{sec:application}, we will focus on the case of two outcomes ($K=2$). We can specify the counterfactual outcome models for each principal stratum. We use $Y_{ij}^1$ and $Y_{ij}^2$ to denote each component outcome, and only always-survivors have well-defined counterfactual outcomes under both treatment and control conditions $\{\boldsymbol{Y}_{ij}(0), \boldsymbol{Y}_{ij}(1)\}$. The protected individuals have only one well-defined counterfactual outcome under the treatment condition, while the never-survivor stratum has a well-defined outcome for neither the treatment nor control condition. %Therefore, for scenarios with a defined QoL outcome, we model each counterfactual outcome as a function of covariates with the cluster-level random effects. 

Denote $\boldsymbol{\alpha}_{g,z}$ as the multivariate regression coefficients, and specifically, $\boldsymbol{\alpha}_{11,1}=(\boldsymbol{\alpha}^{1}_{11,1},\boldsymbol{\alpha}^{2}_{11,1})$, $\boldsymbol{\alpha}_{11,0}=(\boldsymbol{\alpha}^{1}_{11,0},\boldsymbol{\alpha}^{2}_{11,0})$, and $\boldsymbol{\alpha}_{10,1}=(\boldsymbol{\alpha}^{1}_{10,1},\boldsymbol{\alpha}^{2}_{10,1})$. Each of them is a $p$-by-2 matrix of regression coefficients for covariates in the following three groups: always-survivors under the treatment condition ($G_{ij}=11$, $Z_i=1$), always-survivors under the control condition ($G_{ij}=11$, $Z_i=0$) and the protected under treatment ($G_{ij}=10$, $Z_i=1$). Let $\boldsymbol{Y}_{ij}=\left(Y_{ij}^1,Y_{ij}^2\right)^T$ be the bivariate continuous outcome for $\{(i,j) \in S_{ij}(Z_{i})=1\}$.  Let $\boldsymbol{\eta_i}=(\eta_{i}^1,\eta_{i}^2)^T$ be the cluster-level random effects and $\boldsymbol{e}_{ij}=(e_{ij}^1,e_{ij}^2)^T$ be the residual error. Here, all superscripts indicate the dimension of the outcome. The counterfactual outcomes can be modeled through the following set of bivariate linear mixed models (BLMM):
\begin{itemize}
    \item For $G_{ij}=11, Z_i=1$, 
    $$\boldsymbol{Y}_{ij}(1)=
    \begin{pmatrix}   \boldsymbol{X}^T_{ij}\boldsymbol{\alpha}^{1}_{11,1}\\ \boldsymbol{X}^T_{ij}\boldsymbol{\alpha}^{2}_{11,1}
    \end{pmatrix}+ \boldsymbol{\eta}_i+ \boldsymbol{e}_{ij}
  $$
      \item For $G_{ij}=11, Z_i=0$, 
    $$\boldsymbol{Y}_{ij}(0)=
    \begin{pmatrix}    \boldsymbol{X}^T_{ij}\boldsymbol{\alpha}^{1}_{11,0}\\ \boldsymbol{X}^T_{ij}\boldsymbol{\alpha}^{2}_{11,0}
    \end{pmatrix}+ \boldsymbol{\eta}_i+ \boldsymbol{e}_{ij}
  $$
      \item For $G_{ij}=10, Z_i=1$, 
    $$\boldsymbol{Y}_{ij}(1)=
    \begin{pmatrix}    \boldsymbol{X}^T_{ij}\boldsymbol{\alpha}^{1}_{10,1}\\ \boldsymbol{X}^T_{ij}\boldsymbol{\alpha}^{2}_{10,1}
    \end{pmatrix}+ \boldsymbol{\eta}_i+ \boldsymbol{e}_{ij}
  $$
\end{itemize}
where $\boldsymbol{\eta}_i \sim \boldsymbol{\mathcal{N}}(\boldsymbol{0}_{2 \times 1}, \boldsymbol{\Sigma}_{\eta})$, 
$\boldsymbol{\Sigma}_{\eta} = \begin{pmatrix}
    \sigma^{2}_{\eta^1} & \sigma^{2}_{\eta^{12}} \\
    \sigma^{2}_{\eta^{12}} & \sigma^{2}_{\eta^2}
\end{pmatrix}$, 
and $\boldsymbol{e}_{ij} \sim \boldsymbol{\mathcal{N}}(\boldsymbol{0}_{2 \times 1}, \boldsymbol{\Sigma}_{e})$, 
$\boldsymbol{\Sigma}_{e} = \begin{pmatrix}
    \sigma^{2}_{e^1} & \sigma^{2}_{e^{12}} \\
    \sigma^{2}_{e^{12}} & \sigma^{2}_{e^2}
\end{pmatrix}$.

%$\boldsymbol{\eta}_i\sim \boldsymbol{\mathcal{N}}(\boldsymbol{0}_{2 \times 1},\boldsymbol{\Sigma}_{\eta})$, $\boldsymbol{\Sigma}_{\eta}=\begin{pmatrix}
%    {\sigma^{2}_{\eta^1}\ 
%    \sigma^{2}_{\eta^{12}}\\ \sigma^{2}_{\eta^{12}}\ \sigma^{2}_{\eta^2}
%    \end{pmatrix}$ and $\boldsymbol{e}_{ij}\sim \boldsymbol{\mathcal{N}}(\boldsymbol0}_{2 \times 1},\boldsymbol{\Sigma}_{e})$, $\boldsymbol{\Sigma}_{e}=\begin{pmatrix}
% {\sigma^{2}_{e^1}} \ 
%    \sigma^{2}_{e^{12}}\\ \sigma^{2}_{e^{12}} \ {\sigma^{2}_{e^2}}
%    \end{pmatrix}$.

In the set of outcome models, $\sigma^{2}_{\eta^{12}}$ and $\sigma^{2}_{e^{12}}$ denote the covariance for random-effects variance and residual variance. We assume the random effects and the residual errors are mutually independent. For simplicity, we further assume that the potential outcome models share the same random effects across different principal strata and treatment conditions, but one can relax this constraint to model heterogeneous random-effects variance. Following Yang et al.\cite{yang2023power}, with the current parametrization, we can obtain the following four interpretable ICCs among the always-survivors:

\begin{itemize}
    \item Outcome-specific ICCs: $\rho^1=\displaystyle\frac{\sigma^2_{\eta^1}}{\sigma^2_{\eta^1}+\sigma^2_{e^1}}$; $\rho^2=\displaystyle\frac{\sigma^2_{\eta^2}}{\sigma^2_{\eta^2}+\sigma^2_{e^2}}$.
    \item Between-participant, between-outcome ICC: $\rho_1^{12}=\displaystyle\frac{\sigma^2_{\eta^{12}}}{\sqrt{\sigma^2_{\eta^1}+\sigma^2_{e^1}}\sqrt{\sigma^2_{\eta^2}+\sigma^2_{e^2}}}$.   
    \item Within-participant ICC: $\rho_2^{12}=\displaystyle\frac{\sigma^2_{\eta^{12}}+\sigma^2_{e^{12}}}{\sqrt{\sigma^2_{\eta^1}+\sigma^2_{e^1}}\sqrt{\sigma^2_{\eta^2}+\sigma^2_{e^2}}}$.      
\end{itemize}The outcome-specific ICCs capture the similarity between individuals within the same cluster for the two outcomes respectively. The between-participant, between-outcome ICC characterizes the correlation between the two different outcomes between individuals within the same cluster. The within-participant ICC captures the outcome correlation within the same individual. Following the recommendations in the CONSORT extension to CRTs,\cite{campbell2012consort} these ICC parameter estimates can be of interest in CRTs and potentially inform the design of future trials. Therefore, we also study the performance characteristics of the ICC estimators from the proposed joint model.

\subsection{Extension to bivariate binary outcomes}\label{sec:binary}

To model bivariate binary outcomes jointly, we adopt a bivariate Probit random-effects model. For subject $j$ at cluster $i$, let $Y_{ij}^r$ denote the two binary outcomes, where $r=1,2$, and for each outcome, we have:

\begin{equation}\nonumber
P(Y_{ij}^r(z) = 1|\boldsymbol{X}_{ij}^T,\eta_i) =  \Phi^{-1}\left(\boldsymbol{X}_{ij}^T \boldsymbol{a}_{g,z}^r + \eta_i^r \right),
\end{equation}

where $(g, z)$ indexes the principal strata based on combinations of $G_{ij}$ and $Z_i$, and $X_{ij}$ is the observed covariate vector; $\boldsymbol{\Phi}$ is the cumulative distribution function of a standard normal distribution with variance of $\boldsymbol{1}$; the cluster-level random effects $\boldsymbol{\eta}_i = (\eta_i^1, \eta_i^2)^T$ follow a bivariate normal distribution with the same specification as in the BLMM. We can then have the following latent variable specification for each outcome model in different strata. 

\begin{itemize}
    \item For $G_{ij}=11, Z_i=1$, 
    $$Y^1_{ij}(1)
    =\begin{cases}
      1 & \text{if $U^{1}_{11,1}=\boldsymbol{X}^T_{ij}\boldsymbol{a}^{1}_{11,1} + \eta^1_{i}+e^1_{ij} > 0$, }\\
      0 & \text{otherwise}
    \end{cases} \text{and} \ 
    Y^2_{ij}(1)
    =\begin{cases}
      1 & \text{if $U^{2}_{11,1}=\boldsymbol{X}^T_{ij}\boldsymbol{a}^{2}_{11,1} + \eta^2_{i}+e^2_{ij} > 0$, }\\
      0 & \text{otherwise}
    \end{cases}
    $$

    \item For $G_{ij}=11, Z_i=0$, 
    $$Y^1_{ij}(0)
    =\begin{cases}
      1 & \text{if $U^{1}_{11,0}=\boldsymbol{X}^T_{ij}\boldsymbol{a}^{1}_{11,0} + \eta^1_{i}+e^1_{ij} > 0$, }\\
      0 & \text{otherwise}
    \end{cases} \text{and} \ 
    Y^2_{ij}(0)
    =\begin{cases}
      1 & \text{if $U^{2}_{11,0}=\boldsymbol{X}^T_{ij}\boldsymbol{a}^{2}_{11,0} + \eta^2_{i}+e^2_{ij} > 0$, }\\
      0 & \text{otherwise}
    \end{cases}
    $$

    \item For $G_{ij}=10, Z_i=1$, 
    $$Y^1_{ij}(1)
    =\begin{cases}
      1 & \text{if $U^{1}_{10,1}=\boldsymbol{X}^T_{ij}\boldsymbol{a}^{1}_{10,1} + \eta^1_{i}+e^1_{ij} > 0$, }\\
      0 & \text{otherwise}
    \end{cases} \text{and} \ 
    Y^2_{ij}(1)
    =\begin{cases}
      1 & \text{if $U^{2}_{10,1}=\boldsymbol{X}^T_{ij}\boldsymbol{a}^{2}_{10,1} + \eta^2_{i}+e^2_{ij} > 0$, }\\
      0 & \text{otherwise}
    \end{cases}
    $$

\end{itemize}
where $\boldsymbol{\eta}_i \sim \boldsymbol{\mathcal{N}}(\boldsymbol{0}_{2 \times 1}, \boldsymbol{\Sigma}_{\eta})$, 
$\boldsymbol{\Sigma}_{\eta} = \begin{pmatrix}
    \sigma^{2}_{\eta^1} & \sigma^{2}_{\eta^{12}} \\
    \sigma^{2}_{\eta^{12}} & \sigma^{2}_{\eta^2}
\end{pmatrix}$, 
and $\boldsymbol{e}_{ij} \sim \boldsymbol{\mathcal{N}}(\boldsymbol{0}_{2 \times 1}, \boldsymbol{\Sigma}_{e})$, 
$\boldsymbol{\Sigma}_{e} = \begin{pmatrix}
   1 & \rho_e \\
   \rho_e & 1
\end{pmatrix}$.

Here, $\rho_e$ denotes the correlation between the two binary outcomes. In the above latent variable specification, the continuous latent variables $U_{g,z} = \left\{U^{1}_{g,z}, U^{2}_{g,z} \right\}$ follow a normal distribution with unit marginal variances and relate to the observed outcomes $\left\{Y_{ij}^1, Y_{ij}^2 \right\}$ through a threshold at zero. This model for a bivariate binary outcome can be readily extended to accommodate more than two dimensions or to handle mixed outcome types, including combinations of continuous and binary responses.

\section{Bayesian inference}\label{sec:Bayesian}

\subsection{Latent strata membership inference}\label{sec:stratainf}

We use the following data augmentation strategy for imputing the latent principal strata membership,

\begin{itemize}
    \item If an individual is in the treated cluster ($Z_{i}=1$) and does not survive ($S_{ij}=0$), then that individual is a never-survivor ($G_{ij}=00$).
    
    \item If an individual is in the control cluster ($Z_{i}=0$) and survives ($S_{ij}=1$), then that individual is an always-survivor ($G_{ij}=11$).
    
    \item If an individual is in the control cluster ($Z_{i}=0$) and does not survive ($S_{ij}=0$), then that individual is either a never-survivor ($G_{ij}=00$) or protected ($G_{ij}=10$) with the probability of being a never survivor as
    
    $$P(G_{ij}=00|\boldsymbol{X}_{ij}, \boldsymbol{Y}_{ij}, Z_{ij}, S_{ij},\theta)=\frac{P_{ij, G=00}}{P_{ij, G=00}+P_{ij, G=10}}.$$

    \item If one is in the treated cluster ($Z_{i}=1$) and survives ($S_{ij}=1$), then one is either an always survivor ($G_{ij}=11$) or protected ($G_{ij}=10$) with the probability of being an always survivor as
    
    $$P(G_{ij}=11|\boldsymbol{X}_{ij}, \boldsymbol{Y}_{ij}, Z_{ij}, S_{ij},\theta)=\frac{P_{ij, G=11}f_{G=11}(\boldsymbol{Y}_{ij}(1))}{P_{ij, G=11}f_{G=11}(\boldsymbol{Y}_{ij}(1))+P_{ij, G=10}f_{G=10}(\boldsymbol{Y}_{ij}(1))},$$

    where $f_{G=11}(\boldsymbol{Y}_{ij}(1))$ is the likelihood of being in the always-survivor stratum and $f_{G=10}(\boldsymbol{Y}_{ij}(1))$ is the likelihood of being in the protected stratum. Note that in the mixture model, for those in the control arm who do not survive, the full conditional probabilities do not involve potential outcomes since they are not well-defined. On the other hand, for those in the treatment arm and who survive, their potential outcomes are observable such that the full conditional probabilities depend on the outcome likelihood.
\end{itemize}

\subsection{Missingness model and outcome imputation}\label{sec:stratainf}
Following Section \ref{sec:missingnotduetodeath}, the missingness can be handled in two steps. In the first step, 
for participants with both missing survival status and missing QoL outcomes, we impute the survival status using the principal strata membership model described in Section \ref{sec:psmodel}. In the second step, for survived participants with missing outcomes, the arm-specific outcome model for the corresponding stratum can be used to impute the values.

\subsection{Prior specification}\label{sec:prior}
We use a Gibbs sampling approach to draw posterior inferences for our proposed models. A conjugate prior is assigned to each parameter in the potential outcome models and principal strata model. For the potential outcome models, we assign multivariate normal (MVN) priors for the regression coefficients as follows:
$$\boldsymbol{\alpha}_{11,1}\sim MVN(\boldsymbol{a}_{11,1},\boldsymbol{\Sigma}_{11,1})$$
$$\boldsymbol{\alpha}_{11,0}\sim MVN(\boldsymbol{a}_{11,0},\boldsymbol{\Sigma}_{11,0})$$
$$\boldsymbol{\alpha}_{10,1}\sim MVN(\boldsymbol{a}_{10,1},\boldsymbol{\Sigma}_{10,1}).$$
In practice, diffuse priors such as $MVN(\boldsymbol{0}, \text{diag}(1000))$ can be specified. For the random effects variance and residual variance, inverse-Wishart (IW) priors with a degree of freedom $d$ and a scale matrix $\boldsymbol{V}$ can be specified:
$$\boldsymbol{\Sigma_{\eta}}\sim IW(d,\boldsymbol{V_{\eta}})$$
$$\boldsymbol{\Sigma_e}\sim IW(d,\boldsymbol{V_{e}})$$
Non-informative priors can be chosen by setting $d$ to 2 and $\boldsymbol{V}$ to an identity scale matrix. For the principal strata model, we set MVN priors for the regression coefficients with means of $\boldsymbol{b}$ and $\boldsymbol{r}$ and covariance matrices of $\boldsymbol{\Lambda}$ and $\boldsymbol{\Gamma}$ corresponding to the two nested layers of the Probit model:

$$\beta \sim MVN(\boldsymbol{b},\boldsymbol{\Lambda})$$
$$\gamma \sim MVN(\boldsymbol{r},\boldsymbol{\Gamma})$$
The prior for the random effects variance of the principal strata model follows the conjugate inverse Gamma (IG) distribution with shape parameter $g$ and scale parameter $h$:
$$\phi^2 \sim IG(g,h)$$
Diffuse priors can be set by specifying $MVN(\boldsymbol{0}, diag(1000))$, and $IG(0.001,0.001)$.

\subsection{Posteriors inference}\label{sec:posterior}
A Markov Chain Monte Carlo (MCMC) algorithm can used in the posterior inference with random initials. We derived the analytical form of the full conditional distributions and update the model parameters using Gibbs sampling. The algorithm is summarized in Algorithm 1.

\begin{algorithm}\label{algorithm1}
        \hspace*{\algorithmicindent} \textbf{Input: } \text{Set MVN priors for regression coefficients, $\boldsymbol{\alpha}_{11,1}$, $\boldsymbol{\alpha}_{11,0}$, $\boldsymbol{\alpha}_{10,1}$, $\boldsymbol{\beta}$ and $\boldsymbol{\gamma}$; IW priors for $\boldsymbol{\Sigma_{\eta}}$ and $\boldsymbol{\Sigma_{e}}$; an IG prior for $\phi^2$.} \\ \text{Set random initials for all parameters.}  
\begin{algorithmic}[1]
    \For {S iterations }
        \State {Sample $\boldsymbol{\alpha}_{11,1}$, $\boldsymbol{\alpha}_{11,0}$, $\boldsymbol{\alpha}_{10,1}$ from MVN posteriors.
        \State For each cluster, sample $\boldsymbol{\eta_i}$ from a MVN posterior. 
        \State Sample $\boldsymbol{\Sigma_\eta}$ from an IW posterior.
        \State Sample $\boldsymbol{\Sigma_e}$ from an IW posterior.
        \State Sample $\boldsymbol{\beta}$ and $\boldsymbol{\gamma}$ from MVN posteriors.
        \State Sample $\phi^2$ from an IG posterior.
        \State For each cluster, sample $\chi_i$ from a univariate normal posterior.
        \State Update the principal strata membership $G_{ij}$ following the Bayes rule for participants in section 4.1.
        \State Estimate SIACE and SCACE among always survivors($G_{ij}=11$).
        \For {missing outcome not due to death}
            \State impute $\boldsymbol{Y_{ij}}$ using the corresponding BLMM
        \EndFor
        \For {missing outcomes without knowledge of death status} 
            \State Draw $G_{ij}$ (and hence $S_{ij}$) from nested probit model.
            \State If alive ($S_{ij}$=1), impute from BLMM.
        \EndFor
        \State Sample latent variable $Q_{ij}$ and $W_{ij \in (G_{10},G_{11})}$in the nested Probit model from truncated normal distributions.
        %\State *Sample latent variable $\boldsymbol{U}_{11,1}$, $\boldsymbol{U}_{11,0}$ and $\boldsymbol{U}_{10,1}$ in the nested probit outcome model from truncated normal distributions if given bivariate binary outcomes.   
    }
     \EndFor
\end{algorithmic}
    \hspace*{\algorithmicindent} \textbf{Output: } \text{SCACE, SIACE, ICCs} 
    \caption{Pseudo-algorithm for joint modeling of the outcome and principal strata models}
\end{algorithm}

%For a total of S iterations, the algorithm uses Gibbs sampling steps to update outcome regression model parameters, where conjugate priors of multivariate normal and inverse Wishart distributions are specified. As the outcome model also accounts for clustering, it can be updated using the same approach as for $\boldsymbol{\alpha}$. The algorithm further implements a Gibbs sampling step for the principal stratification model where conjugate priors of normal and inverse gamma distributions are used. The clustering effects in the principal strata model denoted as $\chi_i$ can also be updated using a similar approach as for $\boldsymbol{\beta}$ and $\boldsymbol{\gamma}$. Given the regression coefficients, the latent principal strata membership G is imputed following the descriptions from Section 4.1. Then, CSACE and ISACE can be estimated by submitting the always survivors. To fill in and update the missing values, BLMM is used for outcome imputation for individuals with missingness not due to death and for those unknown survival status, outcomes are imputed through BLMM given that they are not truncated by death after drawing their principal strata memberships. Lastly, the latent parameters $Q$ and $W$ in the nested Probit model are updated accordingly from truncated normal distributions. If binary outcomes are given, the additional latent variables, U will be updated using a similar approach as of $Q$ and $W$.

\section{Simulation studies}

\subsection{Simulation design}
We performed simulation studies with a bivariate continuous outcome model under a parallel-arm CRT. We considered four combinations of the average cluster size ($\overline{N}$) and the number of clusters ($n$): $(\overline{N},n)=\{(25,60),(50,60),(25,30),(50,30)\}$.
We simulated two continuous covariates $X_{ij1}$ and $X_{ij2}$ each following a normal or a uniform distribution: $X_{ij1} \sim N(0,100)$ and $X_{ij2} \sim \text{Unif}(-10,10)$. The principal stratification model included an intercept and two covariates, and we considered two sets of regression coefficients $\beta$ and $\gamma$. The first set was $\boldsymbol{\beta}=\{-8.5,0.5,-0.7\}$ and $\boldsymbol{\gamma}=\{-8.8,-0.6,0.4\}$, which resulted in 10\% never survivors, 9\% protected participants and 81\% always survivors. The second set was $\boldsymbol{\beta}=\{-5.5,0.5,-0.7\}$ and $\boldsymbol{\gamma}=\{-5.8,-0.6,0.4\}$, which resulted in 21\% never-survivors, 19\% protected participants, and 60\% always survivors. We used $\boldsymbol{\pi}$ to represent the proportions of participants in different strata.

The outcome models included an intercept, the two simulated covariates, and the cluster size $N_i$. To differentiate SCACE and SIACE, we set coefficients of variability (CV, defined as $\text{CV}=\overline{N}/\sqrt{\text{var}(N_i)}$) as 0.3. For the outcome models, we applied the same set of regression coefficients (including the intercept as the first element) as $\boldsymbol{\alpha}_{11,1}=[\{-13,-0.5,-0.2,0.3\}^T, \{-11,-0.4,0.3,0.3\}^T]$, $\boldsymbol{\alpha}_{11,0}=[\{14,0.5,-0.4,-0.4\}^T, \{12,0.4,0.4,-0.3\}^T]$, and $\boldsymbol{\alpha}_{10,1}=[\{2,2,2,1\}^T, \{-9,-1,0.8,-1\}^T]$. Under 4 different combinations of $(\overline{m},n)$ and 2 sets of $\boldsymbol{\beta}$ and $\boldsymbol{\gamma}$, the outcome-specific SIACE ($\delta^1_I$ and $\delta^2_I$) and SCACE ($\delta^1_C$ and $\delta^2_C$) under the 8 scenarios can be calculated following the formulas in Section \ref{sec:ps}. The SACE results, as well as the model parameters, are summarized in Table \ref{tab:simulation_scenarios}. 
In addition, we also considered principal stratification models that included cluster size as an additional covariate. These corresponding scenarios are summarized in Supplement Table B1. Moreover, we also performed simulation studies where the NMAR assumption was violated by including an additional individual-level covariate in the missingness model. This additional covariate followed Normal distribution with mean 0 and standard deviation 1. Assessed using pseudo-R$^2$, this additional covariate altered from 0.96 and 0.76 to 0.85 and 0.4 respectively. 

The following variance parameters were shared across all scenarios. The random effects variance in the principal strata model was set to $\phi^2=1$ for both layers of the nested Probit model. The random effects variance in the outcome model was set to 1 and 2 for each outcome, respectively, with a covariance of 0.71. The residual variance for the outcome model was set to 5 and 10 for each outcome with a covariance of 3.54. Under this setting, the induced outcome-specific ICCs were 0.167 for both outcomes; the between-participant, between-outcome ICC was 0.08; the within-participant ICC was 0.50. For the missingness model, we defined $\boldsymbol{m}_1=\{10.5,2,-0.1,0.3\}$ and $\boldsymbol{m}_2=\{-0.25,0.5,-0.5,0.2\}$ as the regression coefficients for the first and the second missingness model so that, across all scenarios, roughly 15\% of individuals have both survival status and QoL outcomes missing $(R_{ij}^S=0)$, and roughly 5\% of individuals survived but with missing potential outcomes $(S_{ij}=1, R_{ij}^S=1, R_{ij}^Y=0)$.

For each simulated dataset under each combination of $(\bar{m},n)$, a total of 10,000 MCMC iterations were implemented for the bivariate model with the first 2,500 iterations as burn-in. The average of posterior means, relative bias, coverage, and Monte Carlo errors were calculated across 100 simulated datasets under each scenario for both the bivariate model and the univariate models. For each generated dataset, two univariate models were also implemented for each outcome separately following the strategy in Tong et al. \cite{tong2023bayesian}. The results of the univariate models were compared to those of the bivariate model. Convergence was assessed using trace plots and Geweke's statistics. All analyses were performed using R 4.2.1. The code for the simulation can be found at \url{https://github.com/ttyale/Bivariate-Bayesian-SACE-Missing-Data}.

\begin{table}[]
\centering
\caption{True values of parameters for the simulation scenarios.}
\fontsize{5}{5.5}\selectfont
\label{tab:simulation_scenarios} 

\renewcommand{\arraystretch}{2}
\setlength{\arrayrulewidth}{0.1mm} 
\resizebox{\textwidth}{!}
%\resizebox{\columnwidth}{!}
{%
\begin{tabular}{p{0.3cm}p{0.5cm}p{1.2cm}p{1.2cm}p{2.3cm}p{1.2cm}}
\hline
Scenario & ($\overline{m}$,n)& $\boldsymbol{\beta}$  & $\boldsymbol{\gamma}$ & $\{ \delta_I^1, \delta_I^2, \delta_C^1, \delta_C^2\} $ & $\boldsymbol{\pi} $\\ \hline

I &(25,60) & \multirow{2}{*}{ \scalebox{0.8}{$[-8.5, 0.5, -0.7]^T$ }} & \multirow{2}{*}{\scalebox{0.8}  {$[-8.8, -0.6, 0.4]^T$ }} & \scalebox{0.8}{$[-7.98, -6.78,-9.47, -8.06]^T$} & \multirow{2}{*}{\scalebox{0.8}{$[0.1, 0.09, 0.81]^T$ } }\\

II &(50,60) &  &  & \scalebox{0.8}{$[11.84, 10.22, 8.62, 7.45]^T$}  & \\

III&(25,60)  & \multirow{2}{*}{ \scalebox{0.8}{$[-5.5,0.5,-0.7]^T$ }}& \multirow{2}{*}{\scalebox{0.8}  {$[-5.8, -0.6, 0.4]^T$ }} & \scalebox{0.8}{$[-8.33, -7.18, -9.89, -8.53]^T$}  & \multirow{2}{*}{\scalebox{0.8}{$[0.21, 0.19, 0.60]^T$ } } \\

IV&(50,60)  &  &  & \scalebox{0.8}{$[12.70,10.84,9.45,8.05]^T$}  &  \\

V &(25,30) & \multirow{2}{*}{ \scalebox{0.8}{$[-8.5, 0.5, -0.7]^T$ }} & \multirow{2}{*}{\scalebox{0.8}  {$[-8.8, -0.6, 0.4]^T$ }} & \scalebox{0.8}{$[-10.31,-8.78,-11.65,-9.92]^T$} & \multirow{2}{*}{\scalebox{0.8}{$[0.1, 0.09, 0.81]^T$ } }\\

VI &(50,30) &  &  & \scalebox{0.8}{$[10.73,9.27,7.71,6.68]^T$}  & \\

VII &(25,30)  & \multirow{2}{*}{ \scalebox{0.8}{$[-5.5,0.5,-0.7]^T$ }}& \multirow{2}{*}{\scalebox{0.8}  {$[-5.8, -0.6, 0.4]^T$ }} & \scalebox{0.8}{$[-7.98,-6.89,-9.59,-8.28]^T$}  & \multirow{2}{*}{\scalebox{0.8}{$[0.21, 0.19, 0.60]^T$ } } \\

VIII &(50,30)  &  &  & \scalebox{0.8}{$[5.81,4.94,3.14,2.64]^T$}  &  \\

& \multicolumn{5}{c}{\begin{tabular}[c]{@{}c@{}} Shared in all the scenarios\\ \scalebox{0.8}  {
 
 $\boldsymbol{\alpha}_{11,1}=\begin{bmatrix}-13, -11\\-0.5, -0.4\\-0.2, 0.3\\0.3, 0.3\end{bmatrix},
 
 %$\boldsymbol{\alpha}^{11}_{2,1}=\begin{bmatrix}-11\\-0.4\\0.3\\0.3\end{bmatrix}$,
 
 \boldsymbol{\alpha}_{10,1}=\begin{bmatrix}2 & -9\\2 & -1\\2 & 0.8\\1 & -1\end{bmatrix},
 %$\boldsymbol{\alpha}^{10}_{2,1}=\begin{bmatrix}-9\\-1\\0.8\\-1\end{bmatrix}$,
 
 \boldsymbol{\alpha}_{11,0}=\begin{bmatrix} 14 & 12\\0.5 & 0.4\\-0.4 & 0.4\\-0.4 & -0.3\end{bmatrix},
 %$\boldsymbol{\alpha}^{11}_{2,0}=\begin{bmatrix}12\\0.4\\0.4\\-0.3\end{bmatrix}$}, 
 
 %\scalebox{0.7}
 \boldsymbol{\Sigma}_{\eta}=\begin{bmatrix}1&0.71\\0.71&2\end{bmatrix},
 \boldsymbol{\Sigma}_{e}=\begin{bmatrix}5&3.54 \\3.54&10\end{bmatrix},
 \phi=1$
 %\boldsymbol{\Sigma}_{\eta}=\begin{bmatrix}1&0.71\\0.71&1\end{bmatrix}$
 }

 \end{tabular}} \\ \hline
 
\end{tabular}
}
\end{table}
    
\subsection{Simulation results}

Across all scenarios, the Geweke diagnostic p-values (in Supplement Figure C1-C4) suggest that the ICCs and treatment effect parameters exhibit good MCMC convergence, with p-values generally centered and spread around 0.5 and showing good convergence. The regression coefficients—primarily serving as nuisance parameters—show slightly poorer convergence for certain parameters, but the overall convergence is satisfactory. 

Table \ref{tab:simulation_60} summarizes the simulation results for SACE and ICC parameters for scenarios I-IV with the number of clusters set to 60. The results suggest that the bivariate models have relatively low bias ($<5\%$) and 90\% or above coverage for the posterior means of SIACEs and SCACEs for both outcomes. By contrast, the univariate models have relatively large biases (5\%-30\%) and very low coverage ($<10\%$) for SIACE and SCACE. The univariate models also show larger Monte-Carlo errors in SIACE and SCACE estimates, suggesting that the bivariate model can potentially sharpen the inference compared to the univariate models. In addition, the bivariate model also has relatively low bias and high coverage in the outcome-specific ICCs compared to those in the univariate models. The bivariate model can also estimate the between-participant, between-outcome ICC, and within-participant ICC with high coverage and low bias. 

Table \ref{tab:simulation_30} presents the results for simulation scenarios V-VIII, where the number of clusters is set to 30. The findings for SACE are similar to scenarios I-IV where results of the bivariate model have relatively lower bias, higher coverage, and lower Monte Carlo error compared to those from the univariate model. However, compared to the results of ICC parameters in scenarios I-IV, the bias, coverage, and Monte Carlo error can slightly deteriorate in some scenarios, suggesting that the estimation in ICC parameters can be more challenging as the number of clusters decreases. Still, compared to the univariate models, the ICC results have a much smaller bias and higher coverage. Supplement Tables B2 and B3 presented additional simulation results for scenarios that included cluster sizes as a covariate in the principal stratification model. Findings in these tables were similar where bivariate modeling is clearly more robust, particularly in the estimation of correlation structure. 

Simulation results for when the NMAR assumption was violated are included in Table \ref{tab:sensitivity_60} and Supplement Table D1. The setup of these scenarios assumed missingness mechanism is dependent on unmeasured covariates both layers of the assumption so that the imputation model is misspecified. The results suggest the violation introduces more bias and leads to lower coverage in the treatment effect estimates compared to those in Table \ref{tab:simulation_60}. The impact of violation on the ICC parameters is less substantial. For all paraemeters, the Monte Carlo errors mildly increased but remained low.

\begin{table}
\caption{Simulation results for scenario I-IV with posterior means, bias, coverage, and Monte Carlo error for SIACE, SCACE, and ICC parameters using a bivariate model or univariate outcome models with 60 clusters ($n=60$). Results are average based on 100 simulations.}
\label{tab:simulation_60}
\renewcommand{\arraystretch}{1.4}
\begin{adjustbox}{width=1\textwidth}
\strutlongstacks{T}
\begin{tabular}{ccccccccccc}
\hline
                     &               &            & \multicolumn{4}{c}{Bivariate Model}                             & \multicolumn{4}{c}{Univariate Models}                             \\ \hline
Scenario             & Parameter     & \Longstack{True \\ value} & \Longstack{Posterior \\ mean} & \%Bias & Coverage & \Longstack{Monte-Carlo \\ error} & \Longstack{Posterior \\ mean}& \%Bias & Coverage & \Longstack{Monte-Carlo \\ error} \\ \hline
\multirow{8}{*}{I}   & $\delta_I^1$  & -7.98      & -8.15          & 2.09    & 0.91     & 0.13              & -7.37          & -7.66  & 0.65     & 0.13              \\
                     & $\delta_I^2$  & -6.78      & -6.91          & 1.85  & 0.93     & 0.22              & -7.01          & 3.34   & 0.95     & 0.22              \\
                     & $\delta_C^1$  & -9.47      & -9.79          & 3.35   & 1        & 0.14              & -8.91          & -5.95  & 0.72     & 0.12              \\
                     & $\delta_C^2$  & -8.06      & -8.31          & 3.14   & 1        & 0.21              & -8.28          & 2.73   & 1        & 0.21              \\
                     & $\rho^1$      & 0.17       & 0.17           & -0.77  & 0.98     & 0.0013            & 0.07           & -57.87 & 0.07     & 0.0005            \\
                     & $\rho^2$      & 0.17       & 0.16           & -3.60  & 0.95     & 0.0013            & 0.10           & -38.40 & 0.42     & 0.0007            \\
                     & $\rho_1^{12}$ & 0.08       & 0.08           & -4.11  & 0.96     & 0.0010            & --               &  --     &   --      &   --                \\
                     & $\rho_2^{12}$ & 0.50       & 0.50           & -0.58  & 0.93     & 0.0009            & --               &   --     &    --      &  --                 \\ \hline
\multirow{8}{*}{II}  & $\delta_I^1$  & 11.84      & 11.76          & -0.68  & 0.92     & 0.08              & 13.09          & 10.52  & 0.08     & 0.14              \\
                     & $\delta_I^2$  & 10.22      & 10.14          & -0.71  & 0.93     & 0.16              & 10.00          & -2.14  & 0.98     & 0.28              \\
                     & $\delta_C^1$  & 8.62       & 8.40           & -2.57  & 1        & 0.08              & 9.72           & 12.79  & 0.12     & 0.13              \\
                     & $\delta_C^2$  & 7.45       & 7.26           & -2.53  & 0.99     & 0.15              & 7.24           & -2.77  & 1        & 0.27              \\
                     & $\rho^1$      & 0.17       & 0.16           & -2.58  & 0.96     & 0.0009            & 0.04           & -75.83 & 0        & 0.0002            \\
                     & $\rho^2$      & 0.17       & 0.15           & -7.52  & 0.94     & 0.0009            & 0.04           & -75.67 & 0        & 0.0002            \\
                     & $\rho_1^{12}$ & 0.08       & 0.08           & -9.97  & 0.94     & 0.0007            & --               &  --      &   --       &    --               \\
                     & $\rho_2^{12}$ & 0.50       & 0.49           & -2.41  & 0.96     & 0.0007            & --               &  --      &   --       &    --               \\ \hline
\multirow{8}{*}{III} & $\delta_I^1$  & -8.33      & -8.60          & 3.23   & 0.92     & 0.11              & -6.65          & -20.17 & 0.04     & 0.26              \\
                     & $\delta_I^2$  & -7.18      & -7.39          & 2.90   & 0.94     & 0.21              & -7.51          & 4.61   & 0.97     & 0.33              \\
                     & $\delta_C^1$  & -9.89      & -10.14         & 2.47   & 0.99     & 0.11              & -8.32          & -15.94 & 0.09     & 0.25              \\
                     & $\delta_C^2$  & -8.53      & -8.72          & 2.25   & 1        & 0.21              & -8.87          & 3.97   & 1        & 0.32              \\
                     & $\rho^1$      & 0.17       & 0.16           & -4.91  & 0.98     & 0.0015            & 0.03           & -79.00 & 0        & 0.0004            \\
                     & $\rho^2$      & 0.17       & 0.15           & -12.24 & 0.91     & 0.0014            & 0.06           & -62.51 & 0.12     & 0.0006            \\
                     & $\rho_1^{12}$ & 0.08       & 0.07           & -11.82 & 0.91     & 0.0011            & --               &  --      &  --        &    --               \\
                     & $\rho_2^{12}$ & 0.50       & 0.48           & -4.00  & 0.90     & 0.0012            & --               &  --      &  --        &    --               \\ \hline
\multirow{8}{*}{IV}  & $\delta_I^1$  & 12.70      & 12.65          & -0.37  & 0.93     & 0.09              & 15.72          & 23.81  & 0.01     & 0.33              \\
                     & $\delta_I^2$  & 10.84      & 10.80          & -0.41  & 0.99     & 0.19              & 10.38          & -4.30  & 1        & 0.59              \\
                     & $\delta_C^1$  & 9.45       & 9.32           & -1.35  & 0.99     & 0.09              & 12.21          & 29.16  & 0.01     & 0.31              \\
                     & $\delta_C^2$  & 8.05       & 7.94           & -1.35  & 1        & 0.18              & 7.61           & -5.54  & 1        & 0.59              \\
                     & $\rho^1$      & 0.17       & 0.17           & 4.02   & 0.95     & 0.0010            & 0.02           & -86.49 & 0        & 0.0001            \\
                     & $\rho^2$      & 0.17       & 0.16           & -6.15  & 0.93     & 0.0010            & 0.03           & -83.78 & 0        & 0.0001            \\
                     & $\rho_1^{12}$ & 0.08       & 0.08           & 0.37   & 0.99     & 0.0008            & --               &   --     &   --       &   --                \\
                     & $\rho_2^{12}$ & 0.50       & 0.47           & -4.50  & 0.92     & 0.0011            & --               &  --      &   --       &   --                \\ \hline
\end{tabular}
\end{adjustbox}
\end{table}

\begin{table}[h!]
\caption{Simulation results for Scenario V-VIII with posterior means, bias, coverage, and Monte Carlo error for SIACE, SCACE, and ICC parameters using a bivariate or a univariate outcome model with 30 clusters ($n=30$). Results are average based on 100 simulations.}
\label{tab:simulation_30}
\renewcommand{\arraystretch}{1.4}
\begin{adjustbox}{width=1\textwidth}
\strutlongstacks{T}
\begin{tabular}{ccccccccccc}
\hline
                     &               &            & \multicolumn{4}{c}{Bivariate Model}                             & \multicolumn{4}{c}{Univariate Models}                             \\ \hline
Scenario             & Parameter     & \Longstack{True \\ value} & \Longstack{Posterior \\ mean} & \%Bias & Coverage & \Longstack{Monte-Carlo \\ error} & \Longstack{Posterior \\ mean}& \%Bias & Coverage & \Longstack{Monte-Carlo \\ error} \\ \hline
\multirow{8}{*}{V}    & $\delta_I^1$  & -10.31                                                                 & -10.28                                                                     & -0.33  & 0.93     & 0.29                                                                          & -9.54                                                                      & -7.50  & 0.78     & 0.27                                                                          \\
                      & $\delta_I^2$  & -8.78                                                                  & -8.71                                                                      & -0.72  & 0.93     & 0.47                                                                          & -8.83                                                                      & 0.59   & 0.94     & 0.46                                                                          \\
                      & $\delta_C^1$  & -11.65                                                                 & -11.86                                                                     & 1.84   & 0.99     & 0.28                                                                          & -10.97                                                                     & -5.81  & 0.81     & 0.25                                                                          \\
                      & $\delta_C^2$  & -9.92                                                                  & -10.09                                                                     & 1.70   & 1        & 0.45                                                                          & -10.01                                                                     & 0.94   & 1        & 0.43                                                                          \\
                      & $\rho^1$      & 0.17                                                                   & 0.16                                                                       & -2.51  & 0.96     & 0.0029                                                                        & 0.07                                                                       & -56.45 & 0.42     & 0.0014                                                                        \\
                      & $\rho^2$      & 0.17                                                                   & 0.15                                                                       & -7.40  & 0.89     & 0.0029                                                                        & 0.12                                                                       & -29.61 & 0.95     & 0.0021                                                                        \\
                      & $\rho_1^{12}$ & 0.08                                                                   & 0.07                                                                       & -10.37 & 0.80     & 0.0022                                                                        &    --                                                                        &   --     &    --      &     --                                                                          \\
                      & $\rho_2^{12}$ & 0.50                                                                   & 0.49                                                                       & -2.39  & 0.98     &  0.0021                                                                        &    --                                                                        &   --     &    --      &     --                                                                          \\ \hline
\multirow{8}{*}{VI}   & $\delta_I^1$  & 10.73                                                                  & 10.66                                                                      & -0.71  & 0.94     & 0.19                                                                          & 12.00                                                                      & 11.79  & 0.32     & 0.31                                                                          \\
                      & $\delta_I^2$  & 9.27                                                                   & 9.23                                                                       & -0.43  & 0.91     & 0.35                                                                          & 9.08                                                                       & -1.97  & 0.94     & 0.59                                                                          \\
                      & $\delta_C^1$  & 7.71                                                                   & 7.53                                                                       & -2.34  & 1        & 0.17                                                                          & 8.87                                                                       & 15.02  & 0.4      & 0.28                                                                          \\
                      & $\delta_C^2$  & 6.68                                                                   & 6.53                                                                       & -2.26  & 0.98     & 0.33                                                                          & 6.48                                                                       & -3.00  & 1        & 0.56                                                                          \\
                      & $\rho^1$      & 0.17                                                                   & 0.17                                                                       & 2.87   & 0.94     & 0.0021                                                                        & 0.04                                                                       & -73.26 & 0.08     & 0.0005                                                                        \\
                      & $\rho^2$      & 0.17                                                                   & 0.16                                                                       & -4.76  & 0.96     & 0.0020                                                                        & 0.05                                                                       & -71.60 & 0.09     & 0.0005                                                                        \\
                      & $\rho_1^{12}$ & 0.08                                                                   & 0.08                                                                       & -4.10  & 0.93     & 0.0016                                                                        &                                                                    --        &    --    &    --      &    --                                                                           \\
                      & $\rho_2^{12}$ & 0.50                                                                   & 0.49                                                                       & -2.08  & 0.91     &   0.0018                                                                            &                                                                         --   &   --     &   --       &     --                                                                          \\ \hline
\multirow{8}{*}{VII}  & $\delta_I^1$  & -7.98                                                                  & -8.04                                                                      & 0.76   & 0.96     & 0.26                                                                          & -6.22                                                                      & -22.08 & 0.37     & 0.55                                                                          \\
                      & $\delta_I^2$  & -6.89                                                                  & -6.97                                                                      & 1.13   & 0.92     & 0.45                                                                          & -7.17                                                                      & 4.13  & 0.93     & 0.71                                                                          \\
                      & $\delta_C^1$  & -9.59                                                                  & -9.66                                                                      & 0.75   & 0.99     & 0.25                                                                          & -7.90                                                                      & -17.65 & 0.36     & 0.53                                                                          \\
                      & $\delta_C^2$  & -8.28                                                                  & -8.38                                                                      & 1.25   & 0.99     & 0.44                                                                          & -8.59                                                                      & 3.78   & 0.99     & 0.69                                                                          \\
                      & $\rho^1$      & 0.17                                                                   & 0.17                                                                       & 1.88   & 0.98     & 0.0033                                                                        & 0.03                                                                       & -77.65 & 0.17     & 0.0009                                                                        \\
                      & $\rho^2$      & 0.17                                                                   & 0.14                                                                       & -13.75 & 0.90     & 0.0030                                                                        & 0.07                                                                       & -57.28 & 0.45     & 0.0016                                                                        \\
                      & $\rho_1^{12}$ & 0.08                                                                   & 0.07                                                                       & -16.10 & 0.93     & 0.0023                                                                        &       --                                                                     &    --    &      --    &    --                                                                           \\
                      & $\rho_2^{12}$ & 0.50                                                                   & 0.48                                                                       & -3.28  & 0.95     & 0.0027                                                                        &      --                                                                      &  --      &     --     &     --                                                                          \\ \hline
\multirow{8}{*}{VIII} & $\delta_I^1$  & 5.81                                                                   & 5.71                                                                       & -1.83      & 0.97     & 0.19                                                                          & 8.38                                                                       & 44.17      & 0.07     & 0.59                                                                          \\
                      & $\delta_I^2$  & 4.94                                                                   & 4.72                                                                       & -4.40     & 0.92     & 0.40                                                                          & 4.39                                                                       & -11.01      & 0.95     & 0.99                                                                          \\
                      & $\delta_C^1$  & 3.14                                                                   & 2.97                                                                       & -5.24      & 1        & 0.18                                                                          & 5.51                                                                       & 75.68      & 0.07     & 0.55                                                                          \\
                      & $\delta_C^2$  & 2.64                                                                   & 2.41                                                                       & -8.50      & 0.99     & 0.37                                                                          & 2.14                                                                       & -18.81     & 1        & 0.96                                                                          \\
                      & $\rho^1$      & 0.17                                                                   & 0.17                                                                       & 2.01   & 0.97     & 0.0024                                                                        & 0.02                                                                       & -85.21 & 0.01     & 0.0003                                                                        \\
                      & $\rho^2$      & 0.17                                                                   & 0.15                                                                       & -10.59 & 0.90     & 0.0022                                                                        & 0.03                                                                       & -79.34 & 0.02     & 0.0005                                                                        \\
                      & $\rho_1^{12}$ & 0.08                                                                   & 0.08                                                                       & -10.06 & 0.92     & 0.0017                                                                        &    --                                                                        &  --      &    --      &  --                                                                             \\
                      & $\rho_2^{12}$ & 0.50                                                                   & 0.47                                                                       & -6.89  & 0.89     & 0.0021                                                                        &      --                                                                      &    --    &    --      &    --                                                                           \\ \hline 
\end{tabular}
\end{adjustbox}
\end{table}

\begin{table}[h!]
\small
\center
\caption{\ \ Simulation results for scenario I-IV under the violation of nested missing at random (NMAR) with posterior means, bias, coverage, and Monte Carlo error for ISACE, CSACE, and ICC parameters using a bivariate model with 60 clusters ($n=60$). Pseudo $R^2$ for the two level of missingness are 0.85 and 0.40 respectively. Results are average based on 100 simulations. }
\label{tab:sensitivity_60}
\renewcommand{\arraystretch}{0.99}
\begin{adjustbox}{width=0.85\textwidth}
\strutlongstacks{T}
\begin{tabular}{ccccccc}
\hline
Scenario & Parameter & \Longstack{True \\ value} & \Longstack{Posterior \\ mean} & \%Bias & Coverage & \Longstack{Monte-Carlo \\error} \\ \hline
\multirow{8}{*}{I} & $\delta_I^1$ & -7.89 & -8.25 & 4.47 & 0.83 & 0.14 \\
 & $\delta_I^2$ & -6.69 & -6.91 & 3.17 & 0.92 & 0.23 \\
 & $\delta_C^1$ & -9.46 & -10.15 & 7.27 & 1 & 0.15 \\
 & $\delta_C^2$ & -8.04 & -8.49 & 5.59 & 1 & 0.23 \\
 & $\rho^1$ & 0.17 & 0.17 & 1.48 & 0.94 & 0.0014 \\
 & $\rho^2$ & 0.17 & 0.16 & -1.53 & 0.99 & 0.0014 \\
 & $\rho_1^{12}$ & 0.08 & 0.08 & -1.38 & 0.93 & 0.0011 \\
 & $\rho_2^{12}$ & 0.50 & 0.50 & -0.35 & 0.96 & 0.0010 \\ \hline
\multirow{8}{*}{II} & $\delta_I^1$ & 12.12 & 12.24 & 1.81 & 0.96 & 0.09\\
 & $\delta_I^2$ & 10.46 & 10.55 & 1.81 & 0.92 & 0.17 \\
 & $\delta_C^1$ & 8.96 & 8.87 & -2.38 & 0.99 & 0.09 \\
 & $\delta_C^2$ & 7.75 & 7.67 & -1.33 & 0.99 & 0.16 \\
 & $\rho^1$ & 0.17 & 0.17 & 2.53 & 0.98 & 0.0010 \\
 & $\rho^2$ & 0.17 & 0.16 & -1.81 & 0.94 & 0.0009 \\
 & $\rho_1^{12}$ & 0.08 & 0.08 & -0.17 & 0.94 & 0.0007 \\
 & $\rho_2^{12}$ & 0.49 & 0.50 & -1.06 & 0.93 & 0.0006 \\ \hline
\multirow{8}{*}{III} & $\delta_I^1$ & -6.45 & -6.78 & 5.23 & 0.82 & 0.11 \\
 & $\delta_I^2$ & -5.56 & -5.77 & 3.73 & 0.90 & 0.22 \\
 & $\delta_C^1$ & -8.15 & -8.95 & 10.04 & 1 & 0.10 \\
 & $\delta_C^2$ & -8.02 & -7.60 & 8.31 & 1 & 0.21 \\
 & $\rho^1$ & 0.17 & 0.17 & 0.91 & 0.96 & 0.0015 \\
 & $\rho^2$ & 0.17 & 0.15 & -8.19 & 0.94 & 0.0014 \\
 & $\rho_1^{12}$ & 0.08 & 0.08 & -5.25 & 0.96 & 0.0011 \\
 & $\rho_2^{12}$ & 0.50 & 0.48 & -4.13 & 0.91 & 0.0013 \\ \hline
\multirow{8}{*}{IV} & $\delta_I^1$ & 10.22 & 10.21 & 0.05 & 0.96 & 0.09 \\
 & $\delta_I^2$ & 8.72 & 8.75 & 0.54 & 0.92 & 0.19 \\
 & $\delta_C^1$ & 7.15 & 6.85 & -5.34 & 1 & 0.08 \\
 & $\delta_C^2$ & 6.08 & 5.87 & -4.48 & 0.99 & 0.18 \\
 & $\rho^1$ & 0.17 & 0.16 & 0.59 & 0.94 & 0.0010 \\
 & $\rho^2$ & 0.17 & 0.15 & -6.49 & 0.95 & 0.0011 \\
 & $\rho_1^{12}$ & 0.08 & 0.08 & -4.77 & 0.95 & 0.0008 \\
 & $\rho_2^{12}$ & 0.50 & 0.47 & -5.03 & 0.90 & 0.0010 \\ \hline
\end{tabular}
\end{adjustbox}
\end{table}

\section{The Whole Systems Demonstrator Telecare Questionnaire Study}\label{sec:application}

We illustrate our proposed model with the Whole Systems Demonstrator (WSD) Telecare Questionnaire Study, a two-arm cluster-randomized trial on evaluating the impact of telecare technologies on the QoL of elderly individuals receiving social care in the United Kingdom in 2008 to 2009.\citep{Hirani2013ageing,Hendy2012} In this trial, a total of 1189 participants with an average age of 74 nested in 204 clusters (general practice) were randomized to the intervention and control at the cluster level; 101 clusters were randomized to the treatment and 103 were randomized to the control. The cluster size ranged from 1 to 26, and the average cluster size was 5.8. Participants in the telecare arm received electronic sensors in the homes that provided safety monitoring (e.g., recipient falls, fires in the home), while participants in the usual-care arm received no sensors. 

We study the outcomes of the physical component score (PCS) and mental component score (MCS) assessed by the Short Form 12-item Survey (SF-12) at 12 months post-randomization. \citep{Ware2002}. Each endpoint was a continuous score ranging from 0 to 100, with a higher score indicating better health. The study covariates include several demographics and prognostic variables that were pre-selected for adjustment in the primary analysis, including sex, age, ethnicity, highest level of education, household structure, comorbidity, impairment, and baseline QoL measures. For missing entries in the baseline covariates, we used multiple imputations to impute a single dataset to fill in the missing values. The summary statistics for these variables by treatment arms can be found in Table 4 of Tong et al.\cite{tong2023bayesian}. In this study, various reasons for outcome missingness were documented. A total of 127 participants had missing outcomes due to death or serious deterioration of health, including those unable to continue due to serious illnesses, transferred to long-term nursing care, residential care, family care, or sheltered housing, and dementia with limited mental capacity. Another 62 participants had unobserved outcomes due to non-health-related reasons and were very likely to be alive at the time of assessment. In addition, another 237 participants had both missing outcomes and missing survival status, such as temporary stay in the hospital, and no wishes to share the data or sensor equipment damage.

We applied our bivariate Bayesian model to estimate SIACE, SCACE, and all ICC parameters. We implemented the Gibbs sampling algorithm described in Section 4 by using BLMM for the outcome regression model and a nested Probit model for the principal strata model as previously specified. One MCMC chain of 500,000 iterations was implemented, where the first 100,000 iterations were set as burn-in. We set the initial values for the regression coefficients in the outcome models by using the estimated coefficients from linear mixed-effect models based on complete cases. For the initial principal strata membership, participants with directly identifiable strata memberships were assigned directly, and those with strata
membership not directly identifiable, were randomly assigned to one of the possible strata according to their received treatments and survival status. Diffuse priors $MVN(0,\text{diag}(1000))$ were chosen for $\boldsymbol{\alpha}$, $\boldsymbol{\beta}$, and $\boldsymbol{\gamma}$; covariance hyperparameters $\boldsymbol{\Sigma_{\eta}}$ and $\boldsymbol{\Sigma_e}$ were set to $IW(2,\text{diag}(0.001)$; the prior of $\phi^2$ was set to $IG(0.001,0.001)$. Initial values for latent variables $Q_{ij}$'s and $W_{ij}$'s were generated from truncated normal distributions conditional on initial strata membership assignment as well as initial estimates of $\boldsymbol{X}^T\boldsymbol{\beta}$ and $\boldsymbol{X}^T\boldsymbol{\gamma}$. Model convergence was checked using traceplots and Geweke statistics. We also obtained results using univariate Bayesian models that analyzed the physical and mental health endpoints separately. The univariate model incorporated the missingness model into the univariate model used in Tong et al. \cite{tong2023bayesian} Both SIACE, SCACE, and the outcome-specific ICCs were estimated. In addition, we also fitted linear mixed models to obtain the treatment effects for each endpoint using complete outcomes, though these models do not yield estimates with a clear causal interpretation. All models were fitted using R 4.2.1.

Table \ref{tab:data_example} summarizes the results of the data example. Based on our bivariate outcome model, 74\% participants were identified as always survivors, 9\% participants were protected, and 17\% were never-survivors. The posterior means for SIACE and SCACE for physical health are 0.33 and 0.91, and the 95\% credible intervals are [-0.89, 1.56] and [-0.49, 2.33] respectively. The posterior means for SIACE and SCACE for mental health are -0.22 and 0.51, and their 95\% credible intervals are [-1.43, 1.01] and [-0.84, 1.89].  
Posterior medians are equal to the posterior means, and the credible intervals all included the null. %These findings suggest that the Telecare intervention has no effect on both the physical health and mental health of participants. 
The differences between the two point estimates suggest some mild associations of cluster sizes with the intervention effects, hence implying some degree of informative cluster sizes. In particular, the point estimates of SIACE and SCACE for mental health have opposite signs, suggesting different causal estimands of interest might be in different directions. In addition, we found the posterior means for the outcome-specific ICCs for physical and mental health are 0.02 and 0.04, suggesting a mild level of correlation induced by cluster randomization. The between-outcome with-cluster ICC is 0.02 [0.00,0.07]. The between outcome, within individual ICC is 0.79 [0.77, 0.82], showing a strong correlation between the two endpoints. 
Our results were also compared to univariate models where each of the physical and mental endpoints was analyzed separately. The posterior means and 95\% credible intervals for SIACE, SCACE and outcome-specific ICCs are similar to those obtained from the bivariate model. The estimates from linear mixed models are not qualitatively different, though the point estimates are all slightly smaller than the SIACE and SCACE estimates from the bivariate or univariate models. The differences can be partly driven by the 9\% protected participants, who potentially had worse outcomes than those of the always survivors.

\begin{table}[]
\centering
\caption{Results of SIACE, SCACE, and ICCs in WSD study with the Bayesian multivariate modeling, univariate modeling, and linear mixed-effects model estimates.}
\label{tab:data_example}
\resizebox{\textwidth}{!}{%
\begin{tabular}{cclcclccc}
\hline
\multirow{2}{*}{} & \multicolumn{3}{c}{Bivariate Model} & \multicolumn{3}{c}{Univariate Model} & \multicolumn{2}{c}{Linear Mixed Model} \\ \cline{2-9} 
 & Mean & Median & 95\% CrI & Mean & Median & 95\% CrI & Mean & 95\% CI \\ \hline
SIACE on PCS12 & 0.33 & 0.33 & [-0.89, 1.56] & 0.31 & 0.31 & [-0.86, 1.49] & 0.05 & [-1.02,1.11] \\
SIACE on MCS12 & -0.22 & -0.22 & [-1.43, 1.01] & -0.25 & -0.26 & [-1.47, 0.97] & -0.53 & [-1.50,0.45] \\
SCACE on PCS12 & 0.91 & 0.91 & [-0.49, 2.33] & 0.84 & 0.84 & [-0.52, 2.21] & 0.48 & [-0.65,1.62] \\
SCACE on MCS12 & 0.51 & 0.51 & [-0.84, 1.89] & 0.46 & 0.45 & [-0.89, 1.82] & 0.11 & [-0.94,1,16] \\
$\rho^1$ & 0.02 & 0.02 & [0.00, 0.06] & 0.01 & 0.00 & [0.00, 0.05] & -- & -- \\
$\rho^2$ & 0.04 & 0.04 & [0.01, 0.11] & 0.05 & 0.05 & [0.00, 0.13] & -- & -- \\
$\rho^{12}_1$ & 0.02 & 0.02 & [0.00, 0.07] & -- & -- & -- & -- & -- \\
$\rho^{12}_2$ & 0.79 & 0.79 & [0.77, 0.82] & -- & -- & -- & -- & -- \\ 
$\pi_{00}$  & 0.17 & 0.17 &[0.14, 0.19] &-- & -- & -- & -- & -- \\
$\pi_{10}$  & 0.09 & 0.09 & [0.05, 0.15]& -- & -- & --& -- & -- \\ 
$\pi_{11}$  & 0.74& 0.74 & [0.70, 0.77]& --& --& --& -- & -- \\
\hline
\end{tabular}%
}
\end{table}

\section{Discussion}

The study proposes a Bayesian model for estimating the SACE in cluster-randomized trials, particularly focusing on the challenges presented by different types of missing data. Our primary goal is to estimate the SACE while appropriately accommodating death truncation, missing mortality status, and missing non-mortality outcomes due to reasons other than death. We also leveraged the multi-dimensional nature of the QoL outcomes and presented a bivariate model to better utilize the correlation information between the QoL outcomes. An MCMC algorithm with non-informative priors was proposed for the model estimation. We performed simulation studies to verify the performance of our proposed algorithm in estimating SACE and ICC parameters. The results were also compared to those in univariate models presented in Tong et al. \cite{tong2023bayesian}, where separate models were implemented for each outcome. In general, the proposed bivariate model demonstrates low bias and high coverage for SACE parameters, highlighting its performance in tackling complex missingness and death truncation. Even with non-informative priors, adding additional information from a correlated second outcome can improve the inference. In addition, our model also provides a more accurate estimation for the outcome-specific ICC parameters than the univariate model and can additionally estimate between-participant, between-outcome ICC, and within-participant ICC. We used data from the Whole Systems Demonstrator Telecare Questionnaire Study to exemplify the application of our model and algorithm. Overall, our method shows potential for being broadly used in cluster trials or hierarchical data with complex missing data and death truncation. Our proposed model can be easily extended to a multivariate outcome with 3 or more dimensions or to non-continuous outcome types. It can also serve as a framework for analyzing observation data where the outcome of interest has complex missingness patterns related to an intermediate variable. An additional ignorability assumption on the treatment assignment will be needed in such situations. 

Several structural assumptions or modeling constraints in our model are worth noting. First, 
the homogeneous outcome variance across models can be relaxed to be model-specific.\cite{jo2022handling} The estimation of the heterogeneous variance can, however, be challenging for a small stratum, like the protected participants in the treatment arm, especially when the sample size is small. Second, as been observed previously in Tong et al.\cite{tong2023bayesian} and Wang et al.\cite{Wang2024EM}, the exclusion of the random-effects term ($\chi_i$) in the principal stratification model may have a limited impact on SACE and ICC results. Including the random-effects variance can make the estimation challenging, especially when using a frequentist approach (e.g., the EM algorithm). Third, the monotonicity assumption is generally sound because an intervention is usually designed to be beneficial (in terms of mortality risk) and/or has been through previous pilots. However, monotonicity can be violated in some scenarios, such as when the comparison is made between two active treatments. With our current method of Bayesian modeling, relaxing the monotonicity assumption by allowing for the harmed strata can make the identification of the always survivor stratum become challenging. Alternative model identification strategies (e.g., Hayden et al.\cite{Hayden2005Biometrics} and Tong et al.\cite{tong2025semiparametric}) can be employed in this situation. Fourth, more flexible machine learning and\/ or non-parametric models can be employed for the principal stratification model and the outcome model (especially the latter) to improve efficiency and reduce the bias.\cite{Chen2024AOAS}. They are particularly advantageous if one of the modeling goals is to obtain the treatment effect heterogeneity or identify individual-level treatment effect variation.

\section*{acknowledgement}
 Research in this article is supported by the Patient-Centered Outcomes Research Institute\textsuperscript{\textregistered} (PCORI\textsuperscript{\textregistered} Awards ME-2020C1-19220 to M.O.H. and ME-2020C3-21072 to F.L). F.L. and M.O.H. are funded by the United States National Institutes of Health (NIH), National Heart, Lung, and Blood Institute (grant number R01-HL168202). All statements in this report, including its findings and conclusions, are solely those of the authors and do not necessarily represent the views of the NIH or PCORI\textsuperscript{\textregistered} or its Board of Governors or Methodology Committee. We also acknowledge the generous sharing of WSD data from Drs Shashivadan P Hirani and Stanton P Newman at City, University of London, UK.

\vspace*{1pc}

\noindent {\bf{Conflict of Interest}}

\noindent {\it{The authors have declared no conflict of interest.}}

%\nocite{*}% Show all bib entries - both cited and uncited; comment this line to view only cited bib entries;
\bibliography{template-AMA}%

\end{document}